\documentclass[a4paper,11pt]{article}
\usepackage{pos}

\title{The peculiar new state of the blazar PKS 1510-089}
 \ShortTitle{Peculiar new state of PKS 1510-089}

\author*[a]{H.M. Schutte}
\author[b,a]{M. Zacharias}
\author[a]{M. B\"{o}ttcher}
\author[c]{J. Barnard}

\author[]{for the H.E.S.S. collaboration}
\author[c, d, e, f]{D.A.H.~Buckley}
\author[c]{J.~Cooper}
\author[d, e]{D.~Groenewald}

\affiliation[a]{North-West University, Centre for Space Research, Potchefstroom 2520, South Africa}
\affiliation[b]{Universit\"at Heidelberg, Landessternwarte, K\"onigstuhl 12, D 69117 Heidelberg, Germany}
\affiliation[c]{University of the Free State,  Department of Physics, PO Box 339, Bloemfontein 9300, South Africa}
\affiliation[d]{South African Astronomical Observatory, PO Box 9, Observatory 7935, South Africa}
\affiliation[e]{Southern African Large Telescope Foundation, PO Box 9, Observatory 7935, South Africa}
\affiliation[f]{University of Cape Town, Department of Astronomy, Private Bag X3, Rondebosch 7701, South Africa}

\emailAdd{contact.hess@hess-experiment.eu}

\newcommand{\source}{PKS~1510-089}

\abstract{
Contemporaneous multiwavelength observations with H.E.S.S., SALT, Fermi-LAT, Swift, and ATOM show that the blazar PKS 1510-089 suffered a significant decrease in its optical flux, degree of optical polarization and high-energy gamma-ray ($E > 100$~MeV) flux since July 2021. Meanwhile, the X-ray and very-high-energy gamma-ray (E>100 GeV) fluxes remained steady throughout 2021 and 2022. The degree of optical polarization decreased to about zero in 2022, indicating an unpolarized dominating accretion disk component in the optical-UV domain that is completely diluting the polarized electron synchrotron component. In this proceeding we will discuss, via theoretical SED modeling, possible reasons for this dramatic change in the appearance of this blazar.
}

\ConferenceLogo{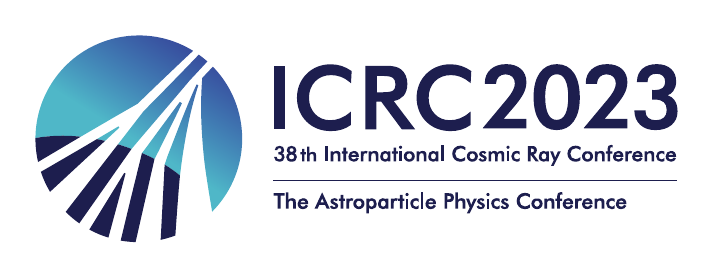}

\FullConference{%
38th International Cosmic Ray Conference (ICRC2023)\\
  26 July - 3 August, 2023\\
  Nagoya, Japan}

\usepackage[nolist]{acronym} 
\usepackage{filecontents} 

\usepackage{floatrow} 

\begin{document}
\maketitle

\section{Introduction}
Blazars are a subtype of active galactic nuclei with jets pointing closely to our \ac{LOS}. They can be modeled with the leptonic model which considers an emitting blob moving along the jet, containing relativistic electrons that produce non-thermal emission through synchrotron radiation and inverse-Compton scattering.
The spectral energy distribution (SED) of PKS 1510 - 089 ($z=0.361$, \cite{burbidgekinman66}) has been described by various models, including scattering of multiple seed photon fields \citep{barnacka+14} or two emission zones \citep{nalewajko+12, prince+19}, one inside the \ac{BLR}, and one within the \ac{DT}, farther along the jet. 
Correlations between the different spectral bands provide inconclusive results \citep{brown13,saito+15,zacharias+19}, motivating models involving more than one emission region. In this proceeding we describe a new peculiar state of \source\ which provides further evidence for multiple emission regions in this blazar \cite{hess23}. In this paper, we provide an overview of the findings presented in our published work \citep{hess23}.
\section{Data Analysis}
\subsection{H.E.S.S.}
Observation times of 50.9h in 2021 and 36.5h in 2022 are accepted from the observation period of 2021 (MJD 59311-59382) and 2022 (MJD 59672-59794) resulting in detection significances of 13.5$\sigma$ and 10.3$\sigma$, respectively. 
The Model analysis chain \citep{denauroisrolland09}  is applied to the data sets using \textsc{very loose} cuts, providing energy thresholds  of \mbox{129 GeV} and 106 GeV, in 2021 and 2022, respectively. 
Light curves and spectra were derived with instrument response functions (IRFs) created with \texttt{Run Wise Simulations} \citep{holler20}. The light curve [Fig.~\ref{fig:mwl_lightcurve}(a)] is not significantly variable. Fig.~\ref{fig:spectra} shows spectra obtained in 2021 and 2022 compared to the detections from \citep{hess13} and \citep{magic18}. 
\subsection{\textit{Fermi}-LAT}
The \ac{Fermi-LAT} data of \source\ were analyzed between 100 MeV and 500 GeV. 
\textsc{FermiTools}\footnote{\url{https://github.com/fermi-lat/Fermitools-conda/wiki}} v.2.2.0 was used with \textsc{P8R3\_SOURCE\_V3}\footnote{\url{http://fermi.gsfc.nasa.gov/ssc/data/analysis/documentation/Cicerone/Cicerone\_LAT\_IRFs/IRF_overview.html}} IRFs including the diffuse emission templates \textsc{gll\_iem\_v07} (galactic) and \textsc{iso\_P8R3\_SOURCE\_V3\_v1} (extragalactic). 
An iterative likelihood procedure is implemented \citep{2018A+C....22....9L} with further details and analysis cuts given in \cite{hess23}.
Light curves are obtained with 3 and 7 days binning from January 2021 to September 2022, see Fig.~\ref{fig:mwl_lightcurve}(b). The \mbox{4FGL-DR3} catalog average is plotted in light grey as a comparison value.
Spectra are derived for 2021 and 2022. In Fig.~\ref{fig:spectra} (top) it can be seen that the spectrum in 2022 is signicantly harder and has significantly reduced flux compared to the spectrum in 2021.
\subsection{\textit{Swift}-XRT}
We analyzed the X-ray data taken with the \ac{Swift} \citep{Gehrels04} in the energy range of 0.3 keV - 10 keV, collected from 2021 to 2022 (ObsIDs 00030797022-00030797027 and 00031173220-00030797029). Observations were conducted in photon counting mode and analyzed with HEASOFT v6.31, with recalibration done through the \verb|xrtpipeline|. Spectral analysis was done with \verb|xspec| \citep{Arnaud96}. The data was binned with 30 counts/bin and fitted with a power-law model with Galactic absorption $N_H = 7.13 \times 10^{20} \; \rm cm^{-2}$ \citep{HI4PI}. The light curve is shown in Fig.~\ref{fig:mwl_lightcurve}(c) and spectra are plotted in Fig.~\ref{fig:spectra} (bottom).
\subsection{\textit{Swift}-UVOT and ATOM}
\textit{Swift}-UVOT \citep{Roming05} data was analyzed using \verb|uvotsource| to obtain fluxes from photons within a region of radius 5". The background within a radius of 10" close to the source, is taken into account. We corrected for dust absorption with extinction $A_{\lambda}$, 
given by a reddening $E(B-V) = 0.0853$ mag \citep{schlaflyfinkbeiner11} and the ratios of $A_{\lambda}/[E(B-V)]$ from \cite{Giommi06}. The \ac{ATOM} \citep{hauser+04} data (in BR filters) was analyzed using the ATOM Data Reduction and Analysis Software and the flux was obtained through differential photometry of 5 custom-calibrated secondary standard stars in the field of view. The extinction correction was applied similarly as for the \textit{Swift}-UVOT data. From the light curves [Fig.~\ref{fig:mwl_lightcurve}(d.) and R-B color in (e.)] it can be seen that there is variability before 2021 July, followed by a decrease and then a steady flux thereafter with 3\% and 2\% fractional variability in the R- and B-bands, respectively. 
Note that the optical photometry coverage is sparse after 2021 July 18, providing only limited information about the flux variability in the optical -- UV regime.
The averaged spectra (excluding the peak in July) is shown in Figs.~\ref{fig:spectra} (bottom) and ~\ref{fig:LEAverage} for 2021 (red) and 2022 (blue). The R-band average from ATOM is taken as an upper limit since 
the ATOM observation windows do not coincide with the \textit{Swift}-UVOT ones.
\subsection{SALT}
\begin{figure*}
\centering
\includegraphics[width=0.96\textwidth]{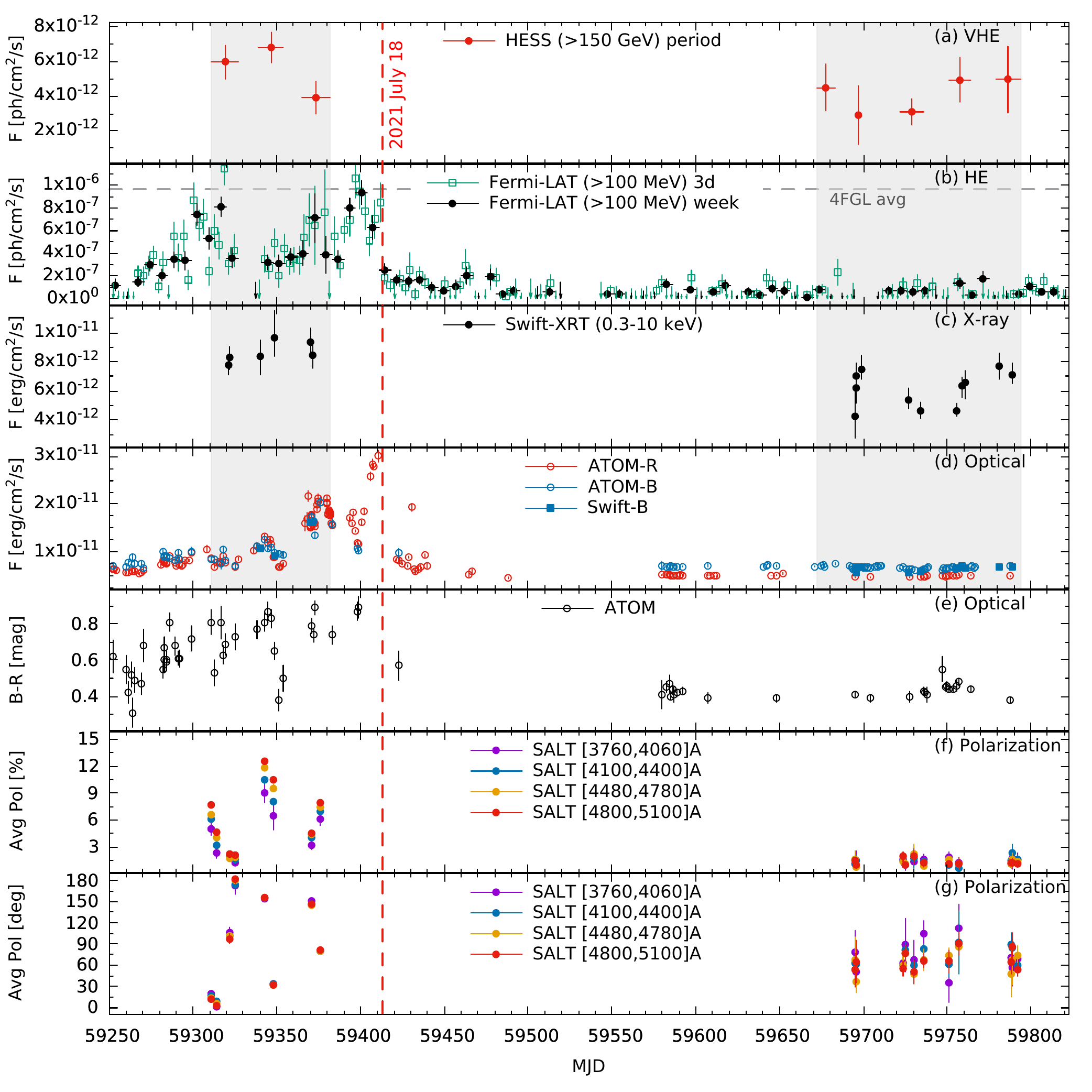}
\caption{Multiwavelength light curves of \source\ in 2021 to 2022 from 
(a) H.E.S.S., (b) \textit{Fermi}-LAT, (c) \ac{Swift}-XRT, as well as (d) ATOM and \textit{Swift}-UVOT. Average spectra (Fig.~\ref{fig:spectra}) are obtained for data in the gray bands. Evolution of B-R color from ATOM (e) and optical polarization from SALT (f and g) is also displayed. This figure is taken from \cite{hess23}.
}
\label{fig:mwl_lightcurve}
\end{figure*}

The \ac{SALT} \citep{2006_SALT_Buckley} observed \source\ in eight observation windows in 2021 and in eleven in 2022, with respective exposure times of 1200~s and 1440~s. The observations made use of the \ac{RSS} \citep{2003SPIE.4841.1463B, 2003SPIE.4841.1634K} (grating PG0900, grating angle 12.875$^{\circ}$ and slit width 1.25"). Observations were conducted in {\sc linear} mode with four wave plate angles available for each observation. The data was analyzed with a modified pySALT/polSALT\footnote{\url{https://github.com/saltastro/polsalt}} \citep{2010SPIE.7737E..25C} pipeline and calibration was done with {\sc IRAF}\footnote{Version 2.16} \citep{cooper22}. In Fig.~\ref{fig:mwl_lightcurve}(f), the degree of polarization is shown in four wavebands to avoid the inclusion of spectral features inbetween these wavebands. The degree of polarization varied between 2.2 $\pm 0.5\%$ and 12.5 $\pm 1.1 \% $ in 2021. In 2022 it stayed below 2\% which is the same as the polarization degree of the comparison star. The degree of polarization detected in 2022 might, therefore, mainly be attributed to the interstellar medium.

\section{Modeling the SED and Optical-UV Polarization}
\begin{figure*}
\includegraphics[width=0.98\textwidth]{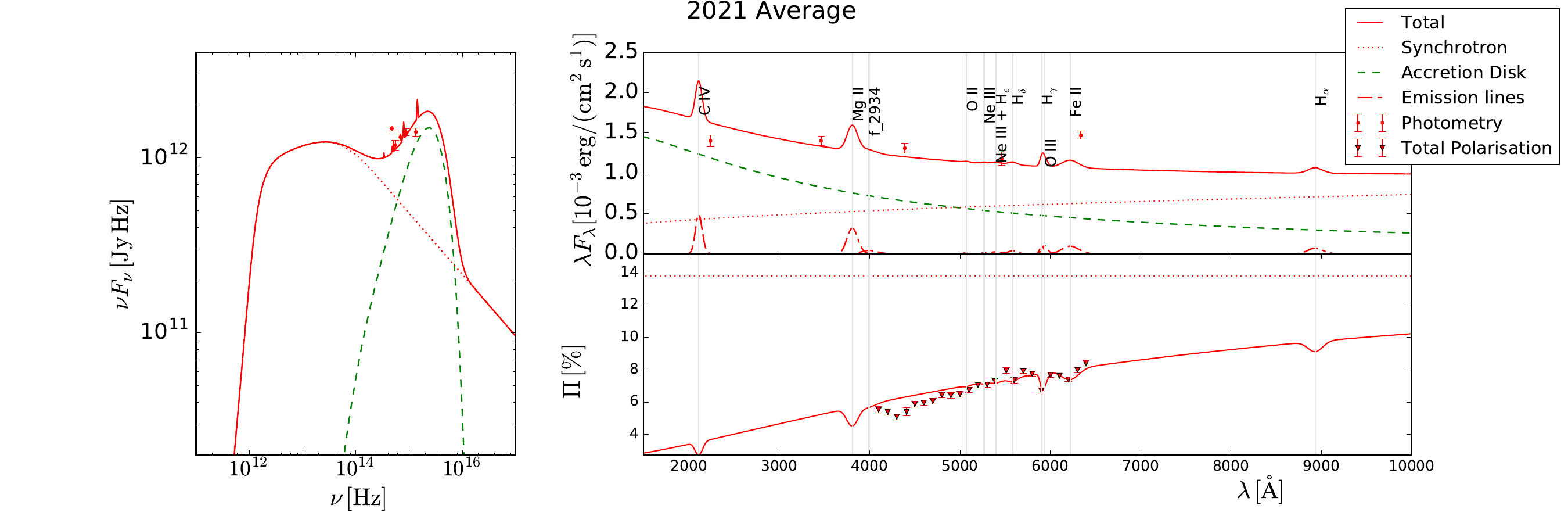}\\
\includegraphics[width=0.98\textwidth]{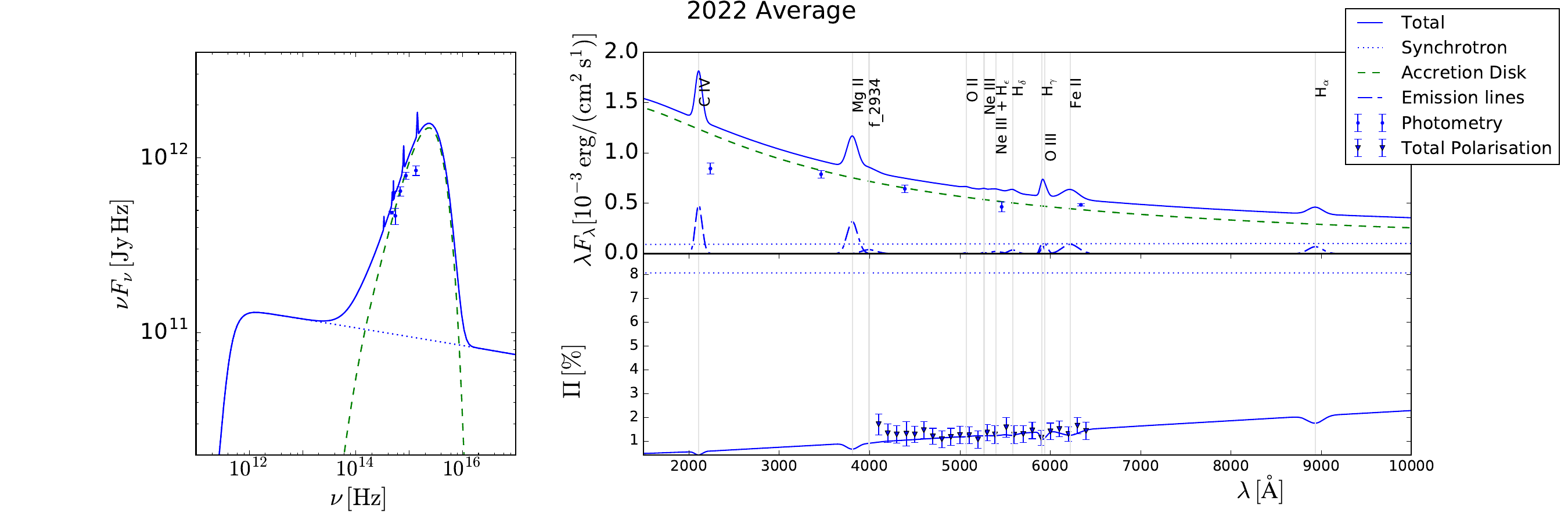}
\caption{Model of \cite{Schutte_2022} applied to the averaged data of 2021 (red) and 2022 (blue). Plots from \cite{hess23}.
}
\label{fig:LEAverage}
\end{figure*}

The low-energy bump in the SED and optical-UV polarization data was fitted with the one-zone model from \cite{Schutte_2022}, as shown in Fig.~\ref{fig:LEAverage}. The electron distribution is defined as a broken (at $\gamma_b$) power-law with exponential cut-off (at $\gamma_c$), within a Lorentz factor range $[\gamma_{\rm min}, \gamma_{\rm max}]$ and spectral indices $p_1$ and $p_2$. Synchrotron self-absorption is taken into account. A geometrically thin, optically thick \ac{AD} \cite{shakurasunyaev73} around a non-rotating black hole is implemented. The black hole mass was taken as $6 \times 10^8 \; \rm M_{\odot}$ (which is within the typcial range \cite{Rakshit2020,Ghisellinietal2010}). The unpolarized \ac{AD} profile remaines constant throughout the 2021-2022 period. The optical-UV polarization degree \cite[Eq.~(6.38)]{1979rpa..book.....R} is dependent on a scaling factor $F_B$ for the ordering of the magnetic field in the range $[0, 1] = $ [completely tangled to perfectly ordered]. The emission lines are coded as Gaussian functions with relative emission line fluxes following \cite{Phillips1978,Malkanetal1986,Isleretal2015}.
By disentangling the polarized synchrotron component from the unpolarized components, we can study the contribution of the synchrotron component in 2021 [Fig.~\ref{fig:LEAverage} (top)].  However, in 2022 [Fig.~\ref{fig:LEAverage} (bottom)], an \ac{AD} and \ac{BLR} (without synchrotron flux) is sufficient to describe the photometry. As the observed polarization degree is equal to the \ac{ISM} contribution, we can only derive an upper limit for the synchrotron polarization degree. Parameters implemented in this model are shown in Tab.~\ref{tab:OptUVpars}.
    
\begin{figure*}[ht]
\floatbox[{\capbeside\thisfloatsetup{capbesideposition={right,bottom},capbesidewidth=7.2cm}}]{figure}[\FBwidth]
{\caption{\textit{Top:} HE and VHE $\gamma$-ray spectra from 2021 (red) and 2022 (blue) compared to the data of \cite{hess13} (dark gray squares) and \cite{magic18} (light gray circles). \textit{Bottom:} One-zone (left) and two-zone (right) leptonic model results \citep{Boettcher+13}. Plots from \cite{hess23}. \\ \label{fig:spectra}}}
{\includegraphics[width=6.8cm]{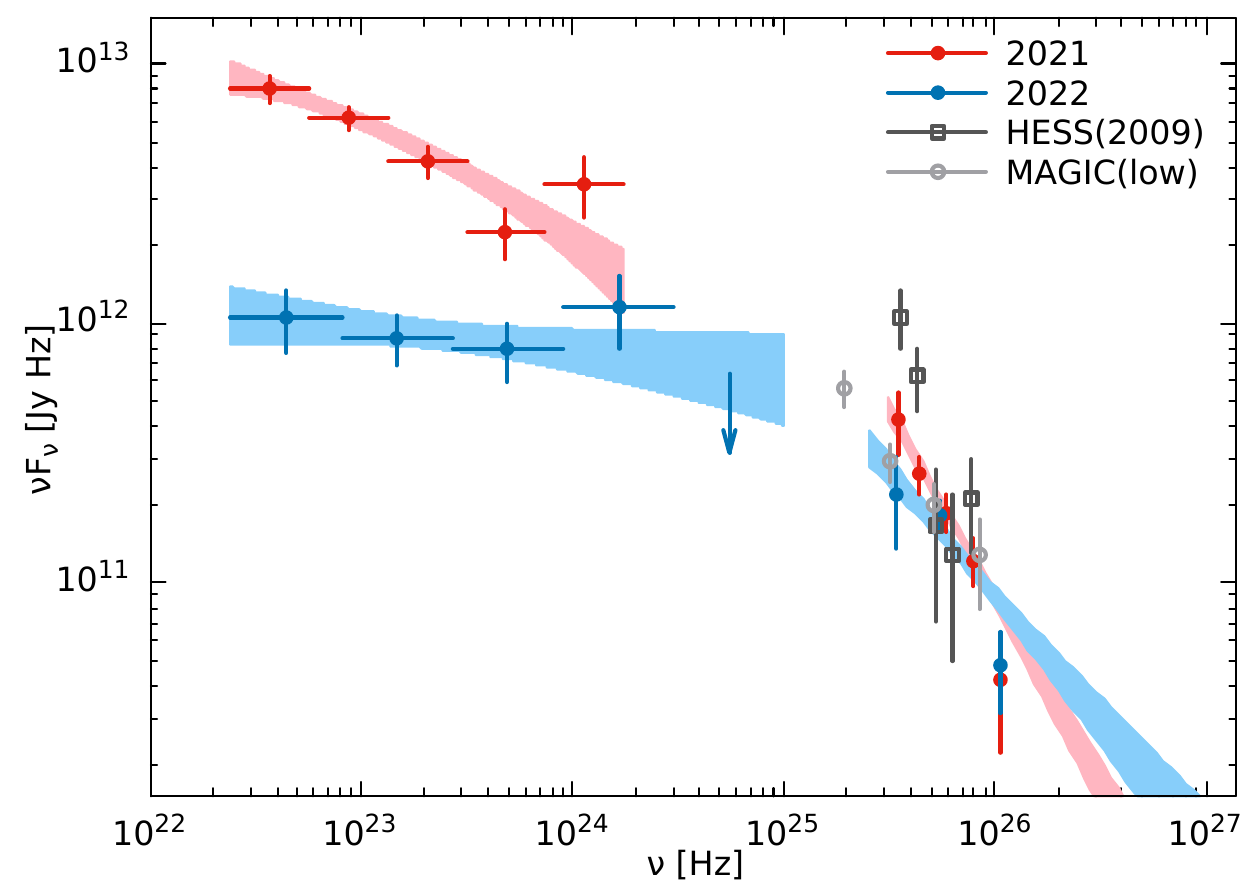}}
\includegraphics[width=0.46\textwidth]{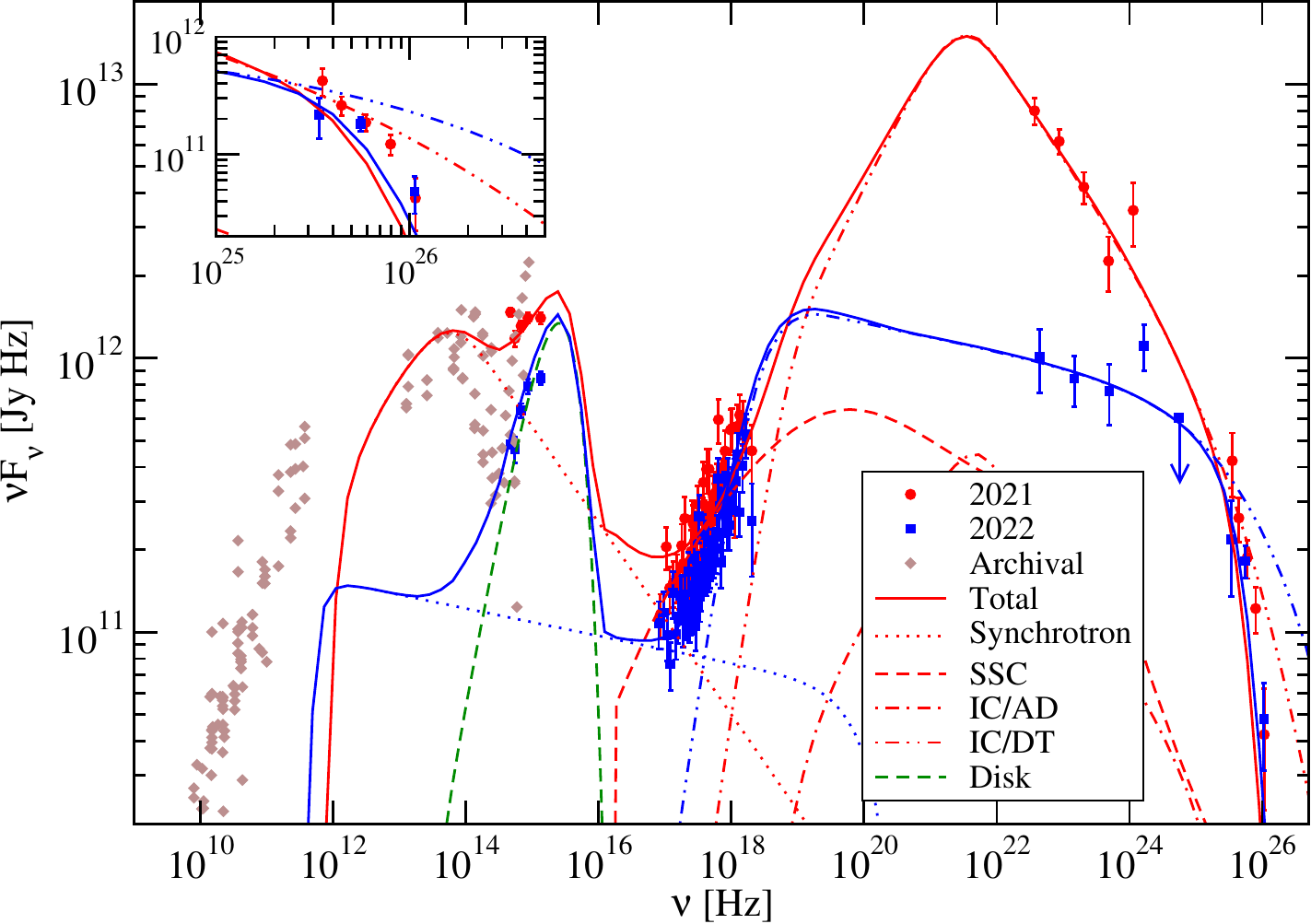} \hspace{0.6cm}
\includegraphics[width=0.46\textwidth]{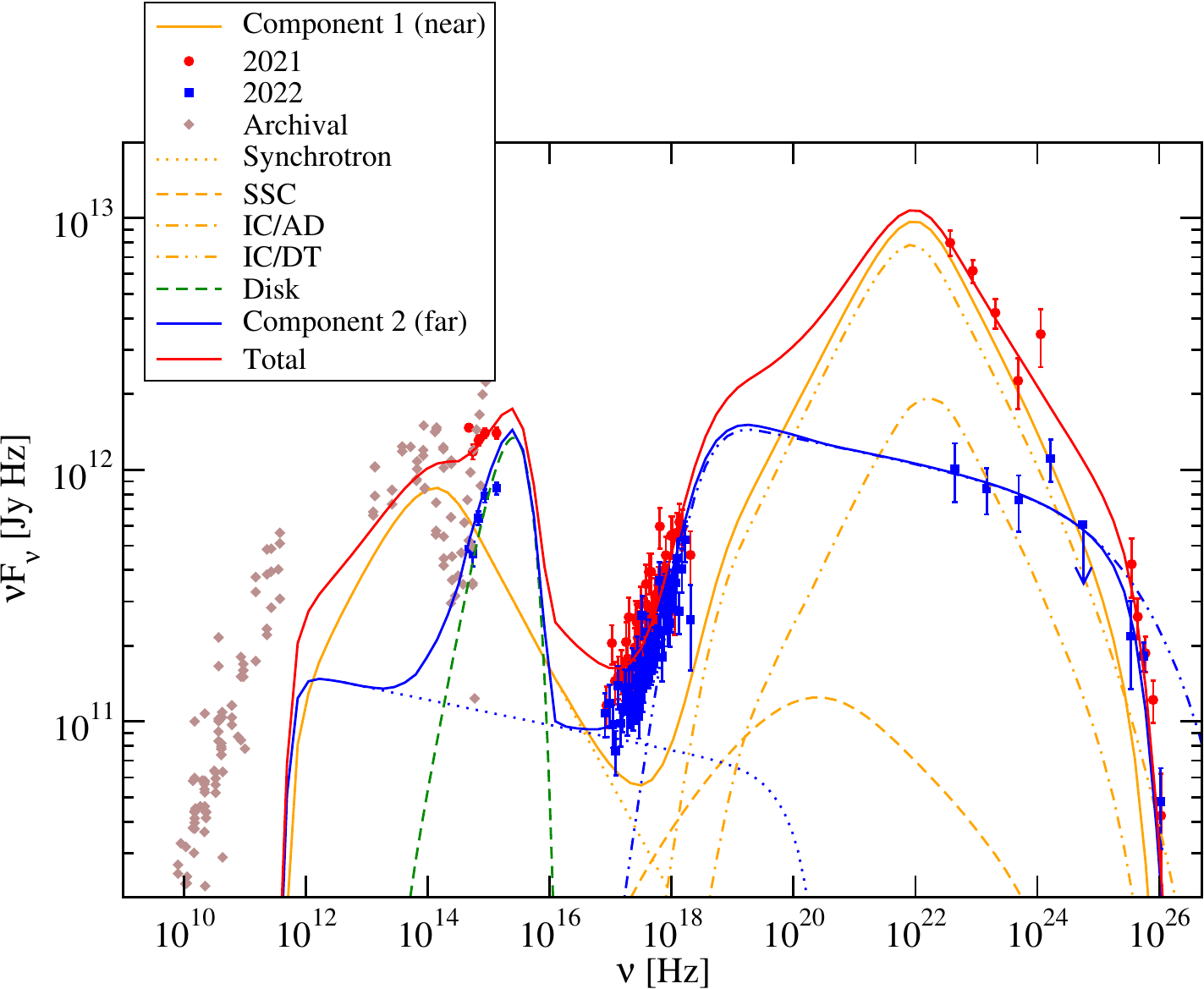}
\end{figure*}

From the observations it is clear that there is a change 
in the optical and \ac{HE} $\gamma$-ray flux state. However, the \ac{VHE} $\gamma$-rays and X-ray fluxes stayed consistent (varying within a factor of 2).
The model of \cite{Boettcher+13} was used to extend our optical-UV results to the broadband \ac{SED}. \ac{SSC} as well as scattering of photon fields from the \ac{AD} and \ac{DT} (modeled as an isotropic black body in the AGN rest frame) determine the inverse-Compton scattering components producing the X- and $\gamma$-ray fluxes in the \ac{SED} [Fig.~\ref{fig:spectra} (bottom)], with corresponding parameters listed in Tab.~\ref{tab:parameters}. The code calculates an equilibrium electron distribution by considering the interplay between injection/acceleration, radiative cooling, and escape. $\gamma\gamma$ absorption by the extragalactic background light is incorporated using the model of \cite{finke+10}.

\begin{table}[ht]
    \centering
    \caption{Parameters included in the code of \citep{Schutte_2022} corresponding to Fig.~\ref{fig:LEAverage}.}\label{tab:OptUVpars}
    \begin{tabular}{l ccccccc}
    \hline
        Date & $n_0$ & $\gamma_b$ & $\gamma_c$ &  $p_1$ &  $p_2$ & $F_B$ & $\chi^2_{pol}/ndf $ \\  \hline
        2021 Average &  $1 \times 10^{47}$ & 569 & $5.0 \times 10^{6}$ & 2.7 & 3.7 & 0.18 & 0.06 \\ 
        2022 Average & $7 \times 10^{49}$ & 30 & $5.0 \times 10^{6}$ & 2.1 & 3.1 & 0.1 & 0.10 \\ \hline
    \end{tabular}
\end{table}

Fig.~\ref{fig:spectra} (bottom) shows an attempt of a one-zone leptonic fit (left).
The switch from the 2021 to the 2022 data set can be achieved by reducing the injection luminosity/acceleration efficiency and adopting a harder injection spectrum. A significant distance from the black hole ($z_0 >> 0.1 \; \rm pc$) is required to minimize any contribution from IC scattering of \ac{AD} photons (IC/AD). Adjusting the fit with a decreased magnetic field ($B \propto z^{-1}$) and larger emission region ($R \propto z_0$) yields a similar fit for the X-ray to \ac{VHE} $\gamma$-ray fluxes and suppresses the synchrotron emission in the radio to X-ray range. In 2021, $z_0$ is not tightly constrained. 

\begin{table}[!ht]
\centering
\caption{\label{tab:parameters} Model parameters corresponding to Fig. \ref{fig:spectra} (bottom). 
}
\begin{tabular}{cccc}
\hline
Parameter [units] & 2021    & 2022   & 2021     \cr
   & single-zone  & single-zone & two-zone \cr
\hline
Kinetic luminosity in jet electrons $L_e$ [erg\,s$^{-1}$]       & $6.2 \times 10^{44}$ & $2.1 \times 10^{44} $ & $2.3 \times 10^{44}$ \cr
$\gamma_{\rm min}$           & 600                  & 30                    & $1.0 \times 10^3$    \cr
$\gamma_{\rm max}$           & $5.0 \times 10^6$    & $1.0 \times 10^6$     & $1.0 \times 10^6$    \cr 
Emission region radius $R$ [cm]                     & $3.0 \times 10^{15}$ & $1.0 \times 10^{16}$  & $5.0 \times 10^{15}$ \cr 
Co-moving magnetic field $B$ [G]                      & 2.0                  & 2.0                   & 2.2                  \cr 
Distance from emission region to black hole $z_0$ [pc]    & 0.1                  & 10                    & 0.06                 \cr 
Bulk Lorentz factor $\Gamma$                     & 20                   & 20                    & 20                   \cr 
Viewing angle in observer's frame $\theta_{\rm obs}$ [deg]     & 2.9                  & 2.9                   & 2.9                  \cr 
Energy density of external blackbody   \cr 
 radiation field $u_{\rm ext}$ [erg\,cm$^{-3}$]    & $1.5 \times 10^{-3}$ & $1.5 \times 10^{-3}$  & $1.5 \times 10^{-3}$ \cr 
Temperature of external blackbody \cr
radiation field $T_{\rm ext}$ [K]            & 100                  & 100                   & 100                  \cr 
Poynting flux power $L_B$ [erg\,s$^{-1}$]        & $6.5 \times 10^{43}$ & $6.0 \times 10^{44}$  & $1.5 \times 10^{44}$ \cr
$L_B/L_e$                    & 0.11 & 2.8                   & 0.66                 \cr
\hline
\end{tabular}
\end{table}
The aforementioned changes required to the one-zone model taking place within a few days in the observer's frame along with a poor fit in the \ac{VHE} domain (see inset in Fig.~\ref{fig:spectra}) render this model implausible.
We therefore suggest a two-zone model [Fig.~\ref{fig:spectra} (bottom, right)]. In 2021 both emission zones are present: 
The primary emission zone, located within the BLR, is responsible for the synchrotron and \ac{HE} $\gamma$-ray radiation and the secondary emission zone, located at $\gtrsim 1$~pc from the black hole, for the X-ray and \ac{VHE} $\gamma$-ray radiation. The X-ray to \ac{VHE} $\gamma$-ray emission remained steady after 2021 July 18, 
while the synchrotron and \ac{HE} radiation were significantly 
reduced. This indicates that the primary emission zone disappeared.

\section{Summary and Conclusions}
This contribution summarizes the outcomes discussed in \citep{hess23}. We presented the observations of a peculiar new state into which \source\ has entered around 2021 July 18. At that time, it experienced a sudden decrease in \ac{HE} $\gamma$-ray flux, optical flux and optical polarization, whereas the \ac{VHE} $\gamma$-ray and X-ray components remained steady into 2022.
The observations support the two-zone leptonic model. Compared to the two-zone interpretation of \cite{nalewajko+12}, our secondary zone requires a softer electron distribution and a slightly higher $\gamma_{\rm min}$ due to differences in the \ac{HE} $\gamma$-ray observations. The new $\gamma$-ray state can be replicated with an IC/CMB model in the kpc-scale jet \cite{Meyer+19}, but this cannot account for the \ac{Swift} spectrum. 
The primary emission zone's disappearance could be due to a significantly weakened inner jet so that substantial radiation can not be produced, or the inner jet has shifted away from the line-of-sight, resulting in reduced Doppler beaming.
%
\begin{acronym}
	\acro{AD}{accretion disk}
	\acro{AGN}{active galactic nuclei}
        \acro{ATOM}{Automatic Telescope for Optical Monitoring}
	\acro{BH}{black hole}
	\acro{BLR}{broad-line region}
        \acro{CMB}{cosmic microwave background}
    \acro{DT}{dusty torus}
	\acro{EC}{external Compton}
	\acro{EBL Abs.}{extragalactic background light absorption}
	\acro{ECD}{external Compton (Disc)}
	\acro{EIC}{external Inverse Compton (BLR)}
	\acro{EM}{electromagnetic}
	\acro{Fermi-LAT}{\textit{Fermi} Large Area Telescope}
	\acro{FSRQ}{flat spectrum radio quasar}
	\acro{GPE}{Gravitational Potential Energy}
	\acro{HBL}{high-frequency peaked BL Lac objects}
	\acro{H.E.S.S.}{\textit{High Energy Stereoscopic System}}
        \acro{HE}{high-energy}
        \acro{ISM}{interstellar medium}
	\acro{IC}{inverse Compton}
	\acro{IR}{infrared}
	\acro{KE}{kinetic energy}
	\acro{LAT}{Large Area Telescope}
	\acro{LBL}{low-frequency peaked BL Lac objects}
	\acro{LCO}{\textit{Las Cumbres Observatory}}
	\acro{LCOGT}{\textit{Las Cumbres Observatory Global Telescope}}
	\acro{LOS}{line of sight}
	\acro{LUVOIR}{\textit{Large UV/Optical/Infrared Surveyor}}
	\acro{MAGIC}{\textit{Major Atmospheric Gamma Imaging Cherenkov Telescope}}
	\acro{NLR}{Narrow Line Region}
	\acro{NRF}{National Research Foundation}
	\acro{QSO}{Quasi Stellar Object}
	\acro{RSS}{Robert Stobie Spectrograph}
	\acro{SAAO}{\textit{South African Astronomical Observatory}}
	\acro{SALT}{\textit{Southern African Large Telescope}}
	\acro{SED}{Spectral Energy Distribution}
	\acro{SMBH}{Super Massive Black Hole}
	\acro{SS}{Shakura Sunyaev}
	\acro{SSA}{Synchrotron Self-Absorption}
	\acro{SSC}{Synchrotron Self-Compton}
	\acro{Swift}{\textit{Neil Gehrels Swift} Observatory}
	\acro{ToO}{Target of Opportunity}
	\acro{UFS}{University of the Freestate}
	\acro{UV}{ultraviolet}
	\acro{VERITAS}{Very Energetic Radiation Imaging Telescope Array System}
        \acro{VHE}{very-high-energy}
	\acro{XRT}{X-ray Telescope }
\end{acronym}

\setlength{\bibsep}{0pt plus 0.3ex} 
\bibliographystyle{JHEP}
\bibliography{references}
%
\end{document}